\documentclass[12pt,a4paper]{article}
\usepackage[latin1]{inputenc}
\usepackage{amsmath}
\usepackage{amsfonts}
\usepackage{amssymb}
\usepackage{graphicx}

\begin{document}

\date{}

\author{ Sunil K. Tripathy\footnote{Department of Physics,
Indira Gandhi Institute of Technology,
Sarang, Dhenkanal, Odisha-759146, INDIA,
tripathy\_ sunil@rediffmail.com},  Dipanjali Behera\footnote{Department of Physics,
Government College of Engineering, Kalahandi,
Bandopala, Risigaon, Bhawanipatna,Kalahandi, Odisha-766002, INDIA,
dipadolly@rediffmail.com} and
Bivudutta Mishra\footnote{Department of Mathematics, Birla Institute of Technology and Science-Pilani,
Hyderabad Campus,Hyderabad-500078, INDIA, bivudutta@yahoo.com} }

\title  {\bf Unified dark fluid in Brans-Dicke theory}

\maketitle

\begin{abstract} Anisotropic dark energy cosmological models are constructed in the frame work of generalised Brans-Dicke theory with a self interacting potential. Unified dark fluid characterized by a linear equation of state is considered as the source of dark energy. Shear scalar is considered to be proportional to the expansion scalar simulating an anisotropic relationship among the directional expansion rates. The dynamics of the universe in presence of unified dark fluid in anisotropic background have been discussed. The presence of evolving scalar field makes it possible to get accelerating phase of expansion even for a linear relationship among the directional Hubble rates. It is found that, the anisotropy in expansion rates does not affect the scalar field, self interacting potential but it controls the non-evolving part of the Brans- Dicke parameter. 
\end{abstract}



\section{Introduction}

Recent observations from distant type Ia supernovae (SNIa) predict that currently the universe is undergoing a state of acceleration \cite{perl98,Reiss98,Reiss04,Knop03}. This intriguing discovery has led to the idea of an exotic form of energy dubbed as dark energy that is responsible for the possible cosmic acceleration at late times. Observations of large scale structure and cosmic microwave background (CMB) also provide strong evidence in favour of dark energy (DE) \cite{Sperg07, Blan06}. The presence of dark energy with a negative pressure is confirmed with additional evidences from observations of X-ray clusters \cite{Allen04}, Baryon Acoustic Oscillations (BAO) \cite{Ein05}, weak lensing \cite{Cont03} and integrated Sache-Wolfe effect \cite{Boughn04, Cole05}. In recent works by Sullivan et al \cite{Sul11} and Suzuki et al. \cite{Suz12} cosmic acceleration with dark energy components has gained much support and a tighter constraint has been put on the dark energy equation of state. The exact nature  of dark energy is not yet known except the fact that dark energy violates the strong energy condition and clusters only at largest accessible scales. Dark energy constitutes the highest contribution to the energy density (68.3 \% dark energy, 26.8\% dark matter and 4.9\% baryonic matter \cite{Ade13,Ade13a, Ade13b}). A simple candidate for dark energy can be a cosmological constant in the classical FRW model with an equation of state equal to $-1$. However, the cosmological constant is entangled with serious puzzles like the fine tuning problem and coincidence problem. Fine tuning problem is concerned with the theoretically predicted value of cosmological constant from quantum field theory  which is larger than the observed value by an order of $10^{123}$. Further it leads to the coincidence problem: "why we are accelerating in the current epoch due that the vacuum and dust energy density are of the same order today?".  Therefore a good number of alternative candidates have been proposed in recent times. Some alternative  candidates for dark energy models are quintessence models \cite{Ratra98}, phantom models \cite{Cald02}, ghost condensate \cite{Piaza04} or k-essence \cite{Chiba00}, holographic DE \cite{Wang08}, agegraphic DE \cite{Cai07, Wei08}, quintom \cite{Eli04, Nojiri05} and so on.  The dark energy provides a negative pressure that generates an anti-gravity effect driving the acceleration. High resolution CMB Radiation anisotropy data from Wilkinson Microwave Anisotropy Probe (WMAP) are in good agreement with the  prediction of the $\Lambda$ dominated cold dark matter model ($\Lambda$CDM) based upon the spatial isotropy and flatness of the universe \cite{Hins09}, \cite{Nolta09}. However, $\Lambda$CDM encounters some anomalous features at large scale. Even though the large scale anomalies in CMB anisotropy are still debatable, WMAP data suggest an asymmetric expansion with one direction expanding differently form the other two transverse directions at equatorial plane \cite{Buiny06} and signal a non-trivial topology of the large scale geometry of the universe \cite{Wat09, Oli04}.

The issue of global anisotropy of the universe can be simply dealt with a simple modification of the FRW model. Recently, some plane symmetric Bianchi-I models or Locally Rotationally Symmetric Bianchi-I (LRSBI) models have been proposed to address the issues related to the smallness in the angular power spectrum of the temperature anisotropy \cite{Campa06,Campa09,Campa07, Grup07}. For a planar symmetry, the universe looks the same from all the points but the points all have a preferred axis. Recent Planck data shows that the primordial power spectrum of curvature perturbation is slightly redshifted from the exact scale invariance \cite{Ade13}. It is obvious from the Planck data that despite the notable success of $\Lambda$CDM model at high multipoles, it does not provide a good fit to the temperature power spectrum at low multipoles \cite{Ade13}. However, it may be noted here that, there still persists uncertainty on these large angle anisotropies and they remain as open problems. LRSBI models are more general than the usual FRW models and are based on exact solutions to the Einstein Field equations with homogeneous but anisotropic flat spatial sections. LRSBI models have also been studied widely, in recent times, in different context \cite{SKT09, SKT10, SKT14,  SKT14a, Koi08, Sharif12}.

Brans-Dicke  (BD) theory is a simple modification of Einstein general relativity where the purely metric coupling of matter with gravity is preserved, thus the universality of free fall (equivalence principle) is ensured \cite{Liu09}. Here, the gravitational constant is replaced with the inverse of a time-dependent scalar field, namely, $\phi(t)=1/8\pi G$, and this scalar field couples to gravity with a coupling constant $\omega$. It passes the experimental tests from solar system \cite{Bert03} and is able to provide an explanation of the accelerated expansion of the universe \cite{Math84}. The theory can also be tested by the observational data coming from CMB and large scale structure \cite{Acq07,Tsuji08, Wu10, Wu13}. Moreover, BD theory arises naturally as the low energy limit of many quantum gravity theories like superstring theory or Kaluza-Klein theory. Since the Brans-Dicke theory has proved to be a better alternative to general relativity and  has a dynamical framework, it evokes  wide interests in the modern cosmology. In view of this, it is worthwhile to discuss dark energy models in this framework. 

In the present work, we have constructed some cosmological models for LRSBI universe in the frame work of BD theory with a self interacting potential and a dynamical BD parameter. Unified dark fluid(UDF), characterized by a linear equation of state, is considered  as the source of dark energy. The paper is organised as follows: In section 2, the basic equations for LRSBI universe are derived. The dynamics of evolution with a Unified dark fluid characterised by a linear equation of state is discussed in Section 3.  We have shown that, a constant deceleration parameter leads to a power law for the BD scalar field. Also, in the work, we concentrate upon a late time dynamics of the universe with accelerated phase of expansion. At late times, the deceleration parameter is believed to be slowly varying or constant. On the other hand, a constant deceleration parameter simulates two kinds of volumetric expansion namely: exponential law and power law.  Cosmological models for exponential expansion and power law expansion are constructed in Section 4 and Section 5 respectively. The dynamics of universe in presence of dark fluid are investigated for respective models. The dynamical BD parameters and self interacting potential for both the models are discussed. Finally, we summarize our results in  Section 6.

\section{Basic Equations}
We consider here the generalized Brans-Dicke (GBD) theory with a self interacting potential. In this GBD theory, the BD parameter is considered as a function of the scalar field $\phi$. The action for GBD theory in Jordan frame is given by \cite{Nord70,Wag70}
\begin{equation}
S=\int d^{4}x \sqrt{-g} [\phi R -\frac{\omega (\phi)}{\phi} \phi ^{,\alpha} \phi_{,\alpha}-V(\phi)+L_{m}],
\end{equation}
where, $\omega (\phi)$ is the modified BD parameter, $V(\phi)$ is the self-interacting potential, $R$ is the scalar curvature and $L_{m}$ is the matter Lagrangian. The unit system we choose here is $8\pi G_{0}=c=1$. Varing the action in (1) with respect to the metric tensor $g_{ij}$ and the scalar field $\phi$, the field equations are obtained as,
\begin{equation}
G_{ij}=\frac{\omega (\phi)}{\phi ^{2}} [\phi_{i} \phi_{j}-\frac{1}{2}g_{ij} \phi_{,\alpha} \phi^{,\alpha}]+\frac{1}{2}[\phi_{,i;j}-g_{ij}\Box \phi]
\end{equation}
\begin{equation}
\Box\phi=\frac{T}{2\omega (\phi)+3}-\frac{2V(\phi)-\phi\frac{\partial V(\phi)}{\partial\phi} }{2\omega(\phi)+3}-\frac{\frac{\partial\omega(\phi)}{\partial\phi} \phi_{,i}\phi^{,i}}{2\omega(\phi)+3}.
\end{equation}

In the above equations, $T=g^{ij} T_{ij}$ is the trace of the energy momentum tensor $T_{ij}$, $\Box$ is the de Alembert's operator. Solar-system experiments predicted a value of the coupling constant as $\omega>40000$ \cite{Bert03}. $\omega$ can be less than 40000 on a cosmological scale \cite{Acq07}. Observational constraints on the Brans-Dicke model were obtained in a flat universe with cosmological constant and cold dark matter using the latest WMAP and SDSS data \cite{Wu10}. Within $2\sigma$ range, the value of $\omega$ satisfies $\omega<-120.0$ or $\omega>97.8$.  In a recent work, the BD parameter is constrained from the combination of observational data of CMB from seven year WMAP, BAO from SDSS, SNIa data from union2 and the X-ray gas mass fraction data from Chandra X-ray observations of the largest relaxed galaxy clusters to be in the range $0.0014 < \frac{1}{\omega}< 0.0024$ or $417 < \omega <714$ \cite{Ala14}. The rate of change of $G$ was constrained to be $-1.75×10^{-12} yr^{-1}<\frac{\dot{G}}{G}<1.05×10^{-12} yr^{-1}$   at $2\sigma$ confidence level in the present epoch\cite{Wu10}.The BD theory reduces to Einstein's general relativity in the limit of a constant scalar field and an infinitely large BD parameter $\omega$. However, this consideration may not hold always good \cite{Sharif12,Romero93,Yadav12}.
\paragraph{}
A plane symmetric LRSBI model is considered through the metric
\begin{equation}
ds^{2}=-dt^{2}+A^{2}dx^{2}+B^{2}(dy^{2}+dz^{2}),
\end{equation}
where $A$ and $B$ are the directional scale factors  and are considered as functions of cosmic time only. The metric corresponds to considering $yz$-plane as the symmetry plane and $x$ as the axis of symmetry. The eccentricity of such a universe is given by $e=\sqrt{1-A^2/B^2}$.  The expansion scalar $\theta$ for this metric is $\theta=\frac{\dot{A}}{A}+2\frac{\dot{B}}{B}$, where, an overhead dot represents ordinary time derivative. Defining the directional Hubble parameters along the axis of symmetry and symmetry plane as $H_1=\frac{\dot{A}}{A}$ and $H_2=\frac{\dot{B}}{B}$, the mean Hubble parameter can be written as  $H=\frac{1}{3}(H_1+2H_2)$ and $\theta=3H$. The scalar expansion can be expressed in terms of the directional Hubble parameters as 

\begin{equation}
\theta=H_1+2H_2.
\end{equation}

The shear scalar for the plane symmetric metric defined in (4) is expressed as 
\begin{equation}
\sigma^2=\frac{1}{2} [\Sigma_i H_i^2-\frac{1}{3} \theta^2 ]=\frac{1}{3} (H_1-H_2 )^2
\end{equation}

The shear scalar may be taken to be proportional to the expansion scalar which envisages a linear relationship between the directional Hubble parameters $ H_1$ and $H_2$ as $H_1=kH_2$. This assumption leads to an anisotropic relation between the directional scale factors $A$ and $B$ as $A=B^k$. Here, $k$ is an arbitrary positive constant that takes care of the anisotropic nature of the model. If $k=1$, the model reduces to be isotropic and  otherwise the model is anisotropic. One can note that such an assumption is not new and is widely used in literature to handle anisotropic models. The mean Hubble parameter can now be expressed as $H=\frac{1}{3}(k+2)H_2$. The average anisotropic parameter $\mathcal{A}=\frac{1}{3}\Sigma \left(\frac{\Delta H_i}{H}\right)^2$ for the model is $\mathcal{A}=2\left(\frac{k-1}{k+2}\right)^2$. Obviously for an isotropic model with $k=1$, $\mathcal{A}$ vanishes and has a finite non zero value for anisotropic models. One should keep it in mind that, the universe is observed to be mostly isotropic and any deviation from isotropic behaviour must be considered as a sort of small perturbation.

The field equations, for a cosmic fluid with energy momentum tensor $T_{ij}=(\rho+p) u_i u_j+pg_{ij}$, now assume the explicit forms
\begin{equation}
9(2k+1)H^2=(k+2)^2 \left[\frac{\rho}{\phi}+\frac{\omega(\phi)}{2}\left(\frac{\dot{\phi}}{\phi}\right)^2-3H\left(\frac{\dot{\phi}}{\phi}\right)+\frac{V(\phi)}{2\phi}\right],
\end{equation}
\begin{equation}
6(k+2)\dot{H}+27 H^2=(k+2)^2 \left[-\frac{p}{\phi}-\frac{\omega(\phi)}{2}\left(\frac{\dot{\phi}}{\phi}\right)^2-\frac{6H}{(k+2)}\left(\frac{\dot{\phi}}{\phi}\right)-\frac{\ddot{\phi}}{\phi}+\frac{V(\phi)}{2\phi}\right],
\end{equation}
\begin{equation}
3(k^2+3k+2)\dot{H}+9(k^2+k+1)H^2=(k+2)^2
\left[-\frac{p}{\phi}-\frac{\omega(\phi)}{2}\left(\frac{\dot{\phi}}{\phi}\right)^2-\frac{3(k+1)H}{(k+2)}\left(\frac{\dot{\phi}}{\phi}\right)-\frac{\ddot{\phi}}{\phi}+\frac{V(\phi)}{2\phi}\right],
\end{equation}
and the Klein-Gordon wave equation for the scalar field,
\begin{equation}
\frac{\ddot{\phi}}{\phi}+3H\frac{\dot{\phi}}{\phi}=\frac{\rho-3p}{2\omega(\phi)+3}-\frac{\frac{\partial\omega(\phi)}{\partial\phi}\dot{\phi}^2}{2\omega(\phi)+3}-\frac{2V(\phi)-\phi\frac{\partial V(\phi)}{\partial\phi}}{2\omega(\phi)+3}
\end{equation}
where, $\rho$ is the dark energy density and $p$ is the dark energy pressure.

Subtracting eqn(9) from eqn(8), we can obtain the evolution equation for the BD scalar field,

\begin{equation}
-\frac{\dot{H}}{H}-3H=\frac{\dot{\phi}}{\phi},
\end{equation}
which can also be expressed  as,
\begin{equation}
(q-2)H=\frac{\dot{\phi}}{\phi},\label{e12}
\end{equation}
where, $q=-1-\frac{\dot{H}}{H^2}$  is the deceleration parameter. It should be mentioned here that a positive deceleration parameter describes a decelerating universe whereas a negative $q$ implies an accelerating one.  Eqn (12) implies that, for a non-static universe $( H\neq 0)$, a constant scalar field will give us a decelerating universe with $q=2$. BD field equations with constant scalar field reduces to the usual Einstein field equations in general relativity. Therefore, one can conclude that in general relativity, accelerating models can not be achieved for LRSBI models by assuming a linear relationship among the directional Hubble rates. This issue has already been investigated earlier \cite{SKT14,SKT13} and similar results have been obtained.  However, in the present work, it is interesting to note that, the presence of an evolving BD field modifies the situation and it is possible to get accelerating models even if the directional Hubble rates are proportional to each other. Again, the behaviour of the BD field is governed by the deceleration parameter and the consequent Hubble rate. For a  constant deceleration parameter the BD field evolves as $\phi \sim a^{q-2}$ or more specifically $\phi \sim (1+z)^{2-q}$, where $a$ is the scale factor and is related to the redshift $z$ as $\frac{1}{a}=1+z$. Here, we consider the scale factor at the present epoch to be $1$. In other words, a constant deceleration parameter favours a power law for the BD scalar field. Moreover, it has become a usual practice, in literature, to use a power law scalar field ($\phi=\phi_0 a^{\alpha}$) to address different issues in cosmology in the framework of BD theory. Also one should keep in mind that Eq. (12) is valid only for an anisotropic model with $k \neq 1$.

\paragraph{}The general expressions for the BD parameter and the self interacting potential can be obtained from the field eqns (7)-(9) as,
\begin{equation}
\omega(\phi)=\left(\frac{\dot{\phi}}{\phi}\right)^{-2}\left[-\frac{\rho+p}{\phi}-\frac{\ddot{\phi}}{\phi}+\frac{3kH}{k+2} \frac{\dot{\phi}}{\phi}-\frac{6\dot{H}}{k+2}-\frac{18(1-k)}{(k+2)^2}H^2\right],
\end{equation}
\begin{equation}
V(\phi)=2\phi\left[\frac{9(2k+1)H^2}{(k+2)^2}-\frac{\rho}{\phi}-\frac{\omega(\phi)}{2}\left(\frac{\dot{\phi}}{\phi}\right)^2+3H\frac{\dot{\phi}}{\phi}\right].
\end{equation}
The behaviour of the BD parameter and the self interacting potential along with the dynamics of the universe can be understood if we know the behaviour of the energy density, pressure and the scale factor of the universe. The scale factor of the universe can be fixed up from the behaviour of the deceleration parameter or the assumed dynamics of the late time accelerated universe. For the pressure and energy density, usually, a barotropic relationship in the form $P=P(\rho)$ , known as equation of state, is assumed. In this sense many equations of state with different mathematical formulations have been proposed in literature to address different issues in cosmology. In the present work, we assume a linear equation of state to handle the issue of dark energy problem in the frame work of generalised BD theory. 

\section{Unified Dark fluid}
A dark fluid model with a linear equation of state was proposed in the spirit of generalized Chaplygin Gas model(GCM) \cite{Bab05,Holman05} after its success in addressing issues related to the late time cosmic acceleration and dark energy problem. Also CGM is known to be quite consistent with observations \cite{Sand04}. Holeman and Naidu in their work in Ref. \cite{Holman05} coined the linear equation of state defining the dark fluid as wet dark fluid (WDF), claiming that such an equation of state is used earlier to treat water and aqueous solution \cite{Tait88, Hayward67}. In UDF, a constant adiabatic sound speed is assumed and the eos is obtained through an integration over the energy density. The integration constant comes out in the process, obviously, has a behaviour similar to the cosmological constant and the  eos has components both from dark matter and dark energy sectors. This is usually referred to as dark degeneracy. 

Unified fluid dark energy is modelled through the equation of state (eos)
\begin{equation}
p=\gamma(\rho-\rho^*)\label{eos},
\end{equation}
where, $\gamma$ and $\rho^*$ are positive constants. This non-homogeneous  linear eos $\eqref{eos}$ provides a description of both hydro-dynamically stable ($\gamma >0 $) and unstable ($\gamma < 0 $) fluids \cite{Bab05}. One can notice here that the UDF eos contains two parts, one behaves as the usual barotropic cosmic fluid and the other behaves as a cosmological constant and unifies the dark energy and dark matter components. The adiabatic speed of sound for this eos  is $C_{s}^2=\gamma$. For stability of a model the adiabatic speed of sound should be $C_{s}^2\geq 0$ and for causality, $C_{s}^2\leq 1$. Hence, $\gamma$ should lie in the range of $0\leq\gamma\leq 1$.  $\gamma=0$, refers to the case of a dark matter and $\gamma=1$ implies a stiff fluid dominated with dark energy ( may be the contribution come from other sources such as a fluid with a bulk viscosity or a cosmological constant). The value of $\gamma$ in between zero and $1$ refers to an exotic cosmic fluid unifying both the dark energy and dark matter and it deals with the dark sector of the universe. However, there are no such constraints for $\rho^*$ and it can be treated as a free parameter. The advantage of the eos  $\eqref{eos}$ is that, dark energy can be described with a positive squared sound speed ( contrary to the need of a negative squared sound speed in phantom energy). In Ref.\cite{Holman05}, Holman and Naidu have claimed that, the WDF model (similar to UDF) is consistent with SNIa observations \cite{Reiss04}, WMAP data \cite{Ben03, Spergel03} and constraints coming from the measurements of matter power spectrum \cite{Tegmark02}. They have shown that, a WDF model with $\gamma=0.316228$ fits well to the observed data. Babichev et al.\cite{Bab05} did not put any sign constraint on the parameters $\gamma$ and $\rho^*$. For different combination of these two parameters they obtained distinctive types of the cosmic evolution scenario such as Big Bang, Big Crunch, Big Rip, anti-Big Rip, solutions with de Sitter attractor and bouncing solutions. They have shown that, for $1+\gamma >0$ and $\gamma \rho^* >0$ the universe may contain either non-phantom or phantom energy whereas for  $1+\gamma >0$ and $\gamma \rho^* <0$ the universe may contain only phantom energy leading to a Big Crunch. On the otherhand, for $1+\gamma <0$ and $\gamma \rho^* <0$, the universe may contain either non phantom or phantom energy whereas for $1+\gamma <0$ and $\gamma \rho^* >0$, the universe may contain only phantom energy leading to a Big Rip in a finite time. The WDF equation of state is considered as a linearised equation of state of any smooth function $p=p(\rho)$ in the vicinity of some local point. UDF dark energy model has generated a considerable research interest in recent times and has been studied widely addressing different issues in relativity and cosmology \cite{Sch06,Mishra13, Chiba97, Chiba98,Balbi07,Xu12,Wang13, Liao12}. 

The parameters of the UDF can be constrained using the observational data on the dark energy equation of state. In the present work, we use the recent observational constraint on dark energy equation of state $\omega_{D}=-1.06^{+0.11}_{-0.13}$ \cite{Kumar14}. The range of allowed values for the parameters $\gamma$ and $\rho^*$ as obtained by using the data of Ref.\cite{Kumar14}  is shown in Figure -1. In the figure, $\gamma$ is restricted within the range $0 \le \gamma \le 1$ basing upon the stability and causality of the model which keeps the parameter $\rho^*$ in the positive domain for negative $\omega_{D}$. In a recent work, Liao et al. \cite{Liao12} have constrained the parameters of a unified dark fluid described through a two parameter affine linear equation of state similar to the one discussed in this work using the Hubble parameter data $H(z)$, type Ia Supernovae data from Union 2 datasets, Baryon Acoustics Oscillations observations from Sloan Digital Sky Survey and the CMB radiation data from WMAP. They have constrained the parameter $\gamma$ to be $0.00172^{+0.00392}_{-0.00479}$ in $1\sigma$ for a flat universe and $0.00242^{+0.00787}_{-0.00775}$ in $1\sigma$ for a non-flat universe. In another work, Xu et al. \cite{Xu12} constrained this parameter to be $0.000487^{+0.000117}_{-0.000487}$ in $1\sigma$ confidence. So far, it is believed that a low value of $\gamma$  much less than $1$ fits the observational data well.

\begin{figure}[h!]
\begin{center}
\includegraphics[width=1\textwidth]{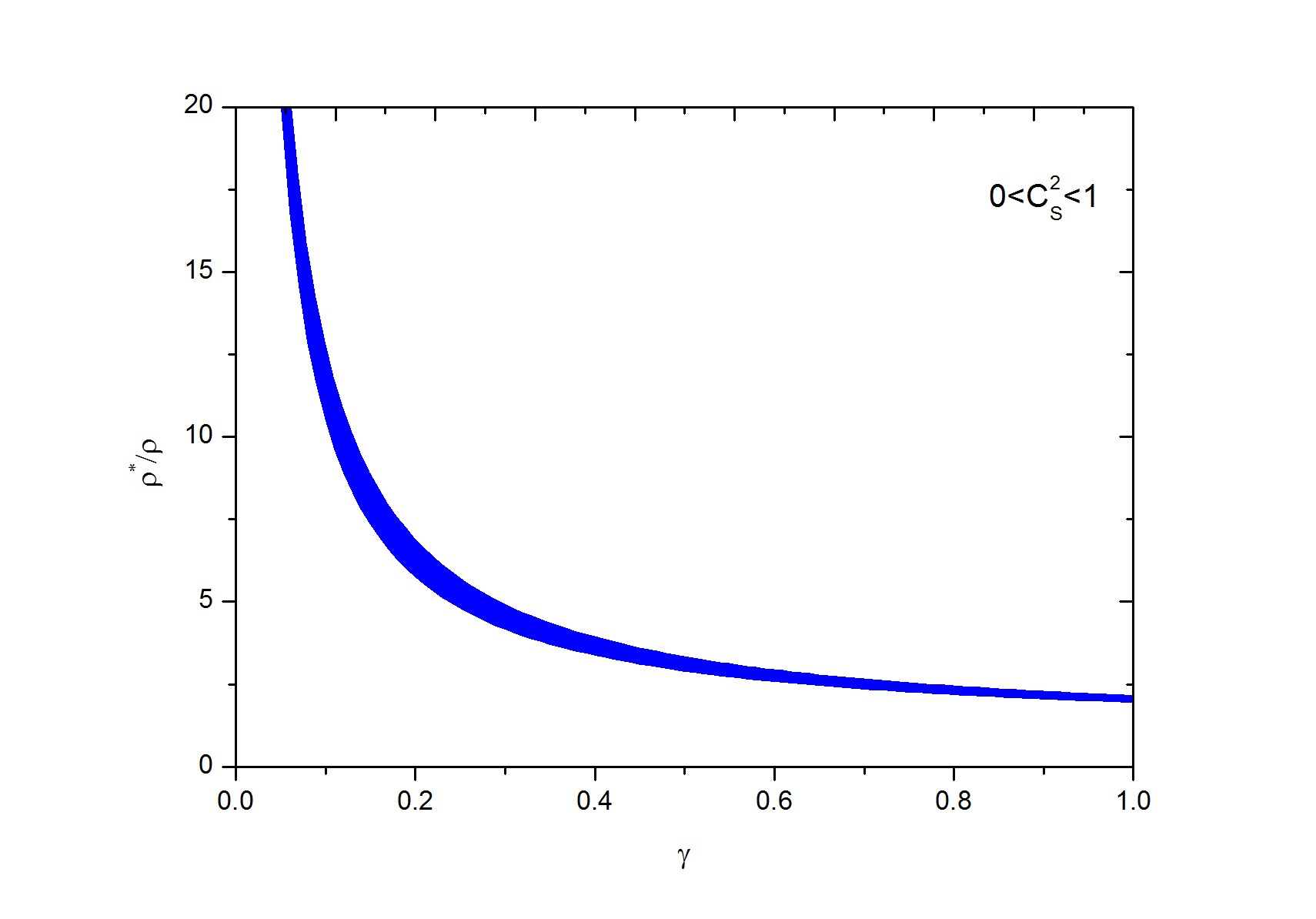}
\caption{Observational constraints on the UDF parameters.}
\end{center}
\end{figure}

The energy conservation equation for matter field is given by
\begin{equation}
\dot{\rho}+3H(\rho+p)=0.
\end{equation} 
For unified dark fluid eos, (16) can be integrated to get 
\begin{equation}
\rho=\rho_{\Lambda}+ \rho_{\gamma}a^{-3(1+\gamma)}
\end{equation}
where $\rho_{\Lambda}=\frac{\gamma\rho^*}{1+\gamma}$ and $\rho_{\gamma}=(\rho_0-\rho_{\Lambda})$. $a=(AB^2 )^{\frac{1}{3}}$ is the average radius scale factor of the universe. $\rho_0$ is the dark energy density at the present epoch. Since $\gamma$ and $\rho^*$ are positive, $\rho_{\Lambda}$ is positive varying between  $0$ and $\frac{\rho^*}{2}$ for $\gamma=0$, $\gamma=1$ respectively.  Depending upon the relative values of $\rho_0$ and $\rho_{\Lambda}$, $\rho_{\gamma}$ can either be positive or negative. It is interesting to note that, the dark energy density has two parts: one behaves like a cosmological constant and the other part dynamically evolves with the cosmic expansion.

The dark energy pressure can be expressed as 

\begin{equation}
p=-\rho_{\Lambda}+ \gamma\rho_{\gamma}a^{-3(1+\gamma)},
\end{equation}
so that the equation of state parameter $\omega_{D}=\frac{p}{\rho}$ becomes

\begin{equation}
\omega_{D}=-1+\frac{1+\gamma}{1+\left(\frac{\rho_{\Lambda}}{\rho_{\gamma}}\right) a^{3(1+\gamma)}}.
\end{equation}

The dynamical evolution of the DE eos can  also be assessed from

\begin{equation}
\omega_{D}=-1+\frac{1+\gamma}{1+\left(\frac{\rho_{\Lambda}}{\rho_{\gamma}}\right) (1+z)^{-3(1+\gamma)}}.
\end{equation}

The dark energy pressure and the DE equation of state parameter also have two parts each, one corresponds to the usual cosmological constant and the second part evolves dynamically with cosmic expansion. In Figure-2, the dynamical evolution of the DE eos parameter is shown as a function of redshift for three representative values of the ratio $\frac{\rho_{\Lambda}}{\rho_{\gamma}}=20,30$ and $50$ corresponding to $\omega_{D}= -0.937, -0.958$ and $-0.974$ at the present epoch. $\gamma$ is chosen to be $0.316$. $\omega_{D}$ dynamically evolves from $\gamma$ at early epoch to $-1$ at late times of evolution. In the intermediate time zone, the behaviour of the DE eos is the same for all the choices of $\frac{\rho_{\Lambda}}{\rho_{\gamma}}$, except the fact that, with increase in the value of the ratio, $\omega_{D}$ becomes less negative. In Figure-3, the DE eos is plotted as a function of redshift with $\gamma=0.316$ for three negative values of the ratio $\frac{\rho_{\Lambda}}{\rho_{\gamma}}=-8,-20$ and $-50$ corresponding to $\omega_{D}= -1.19, -1.07$ and $-1.03$ at the present epoch. The DE eos evolves in the phantom region and increases with the cosmic expansion to behave like a cosmological constant.

\begin{figure}[h!]
\begin{center}
\includegraphics[width=1\textwidth]{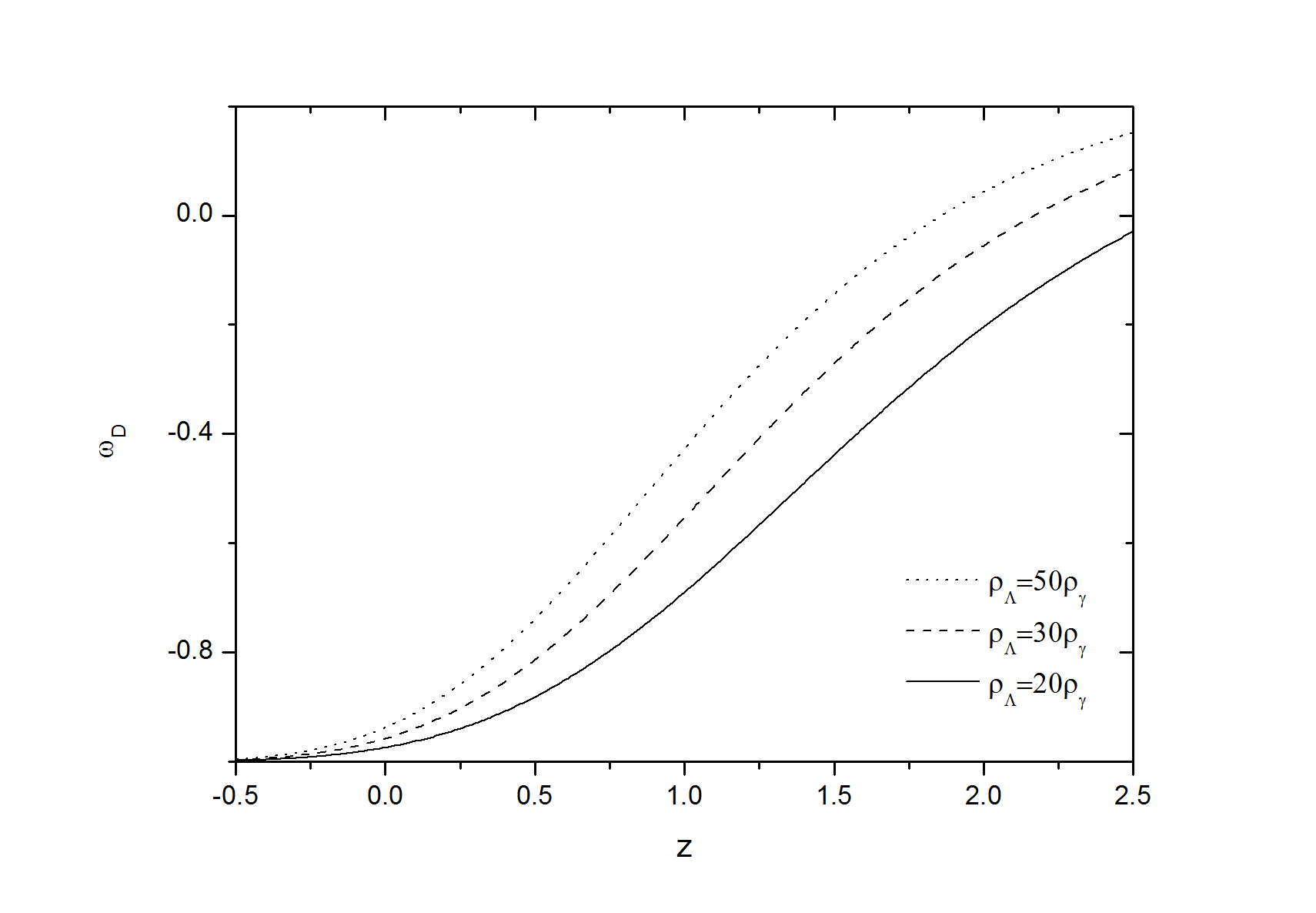}
\caption{Dark energy equation of state as a function of redshift for three positive values of the ratio $\frac{\rho_{\Lambda}}{\rho_{\gamma}}$.  $\gamma$ is taken to be 0.316.}
\end{center}
\end{figure}

\begin{figure}[h!]
\begin{center}
\includegraphics[width=1\textwidth]{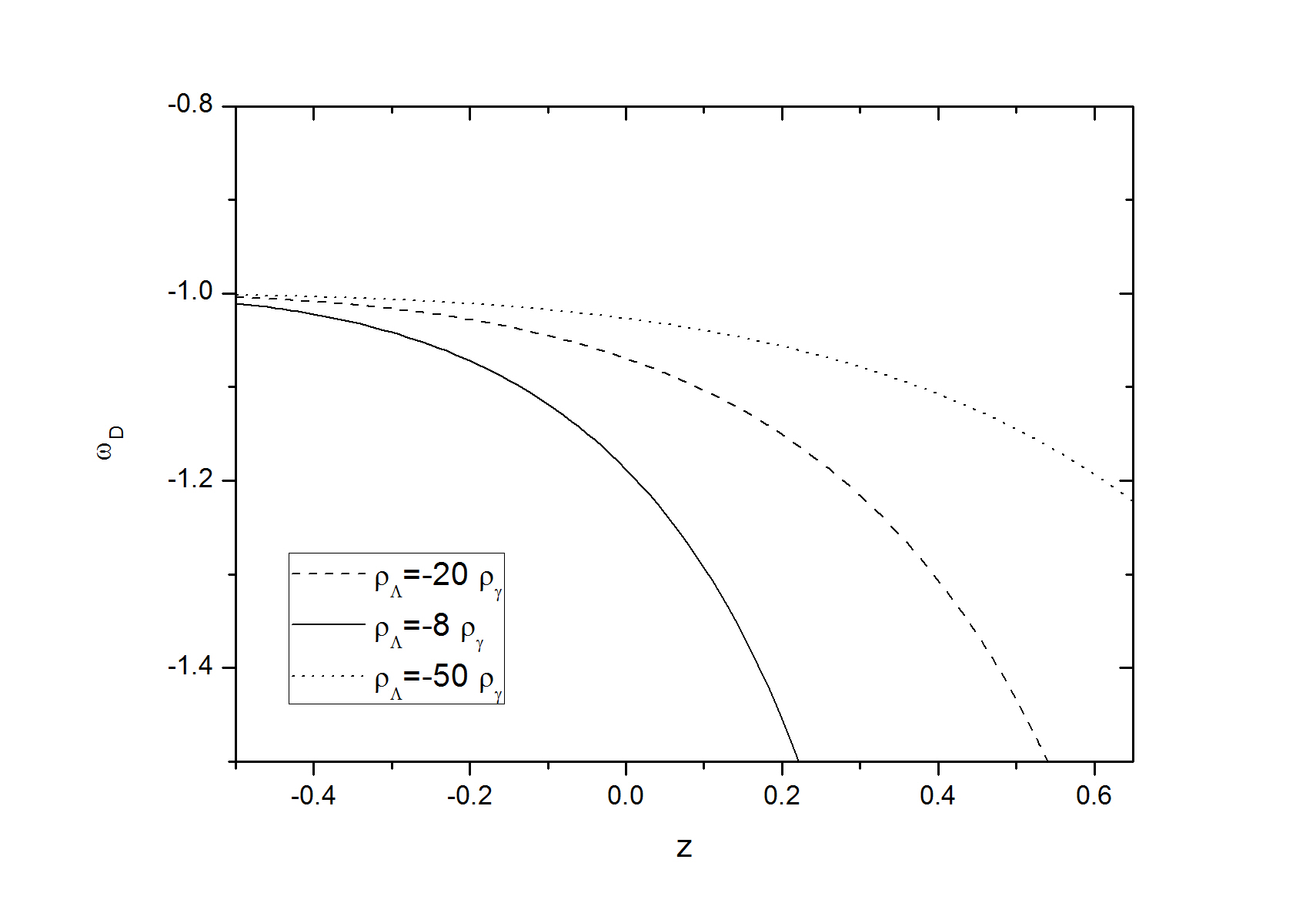}
\caption{Dark energy equation of state as a function of redshift for three negative values of the ratio $\frac{\rho_{\Lambda}}{\rho_{\gamma}}$.  $\gamma$ is taken to be 0.316.}
\end{center}
\end{figure}

Deceleration parameter $q=-\frac{\ddot{a}}{aH^2}$and jerk parameter $j=\frac{\dddot{a}}{aH^3}$ are considered as important quantities in the description of the dynamics of universe.  The observational constraints as set upon these parameters in the present epoch from type Ia supernova and X-ray cluster gas mass fraction measurements are $q_0=-0.81\pm 0.14$ and $j_0=2.16\pm_{-0.76}^{+0.81}$ \cite{Rapetti07}. In a recent work, the deceleration parameter is constrained from  $H(z)$ and SNIa data to be $q=-0.34\pm 0.05$ \cite{Kumar12}. Experimentally it is challenging to measure the deceleration parameter and jerk parameter and one needs to observe objects of red shift $z\ge 1$.  In attempts to investigate the accelerated expansion of the universe, the sign and behaviour of these parameters have been considered in different manner in different works. The time variation of the deceleration parameter is under debate eventhough in certain models, a time varying $q$ leads to a  cosmic transit from early deceleration to late time acceleration \cite{Yadav13,Adhav11,Akarsu12,Pradhan07}. However, at late of time of cosmic expansion, the deceleration parameter is believed to vary slowly with time or becomes a constant. A constant deceleration parameter leads to two different volumetric expansion of the universe namely the power law expansion and exponential expansion. In a model with exponential expansion, the radius scale factor increases exponentially with time leading to a constant Hubble rate. Whereas in a model with power law expansion of the volume scale factor, the scale factor can be expressed as a cosmic time raised with some positive power. The Hubble parameter for such a power law model behaves reciprocally to the cosmic time. In the present work, we are interested in models describing a late time universe with predicted cosmic acceleration and therefore we will consider the exponential and power law expansion of the scale factor corresponding to a constant and variable (decaying) mean Hubble rate i.e $H=H_0$ and $H=\frac{m}{t}$, where $H_0$ and $m$ are positive constants. It is worth to mention here that, the choice of a constant deceleration parameter can not provide a time dependent cosmic transition from a deceleration phase in the past to an accelerated phase at late times.

\section{Exponential Model}
In this kind of volumetric expansion, the Hubble rate is a constant quantity i.e.  $H=H_0$=constant and the scale factor is given by $a=e^{H_0 (t-t_0)}$ and it describes a de Sitter type universe. $t_0$ is the cosmic time in the present epoch. The directional scale factors along the longitudinal and  transverse directions are $A= e^{\frac{3kH_0 (t-t_0)}{(k+2)}}$ and $B= e^{\frac{3H_0 (t-t_0)}{(k+2)}}$. The deceleration parameter and jerk parameter for this choice of the Hubble rate, are $q=-1$ and  $j=1$. The directional deceleration parameters $q_x, q_y$ and $q_z$ are the same as that of the mean deceleration parameter $q$. 

\paragraph{}Integration of (12) yields for an exponential scale factor, 

\begin{equation}
\phi=\phi_0 e^{-3H_0(t-t_0)},
\end{equation}

where, $\phi_0$ is the value of the scalar field in the present epoch. In terms of the scale factor and redshift $z$ , we can express the scalar field  respectively as $\phi=\phi_0 a^{-3}$ and $\phi=\phi_0 (1+z)^{3}$, where we have used the fact $\frac{1}{a}=1+z$. In Figure-4, the evolution of the BD scalar field is plotted as a function of redshift. The scalar field decreases exponentially from a large value at the early epoch to vanish at late times of cosmic evolution.

\begin{figure}[h!]
\begin{center}
\includegraphics[width=1\textwidth]{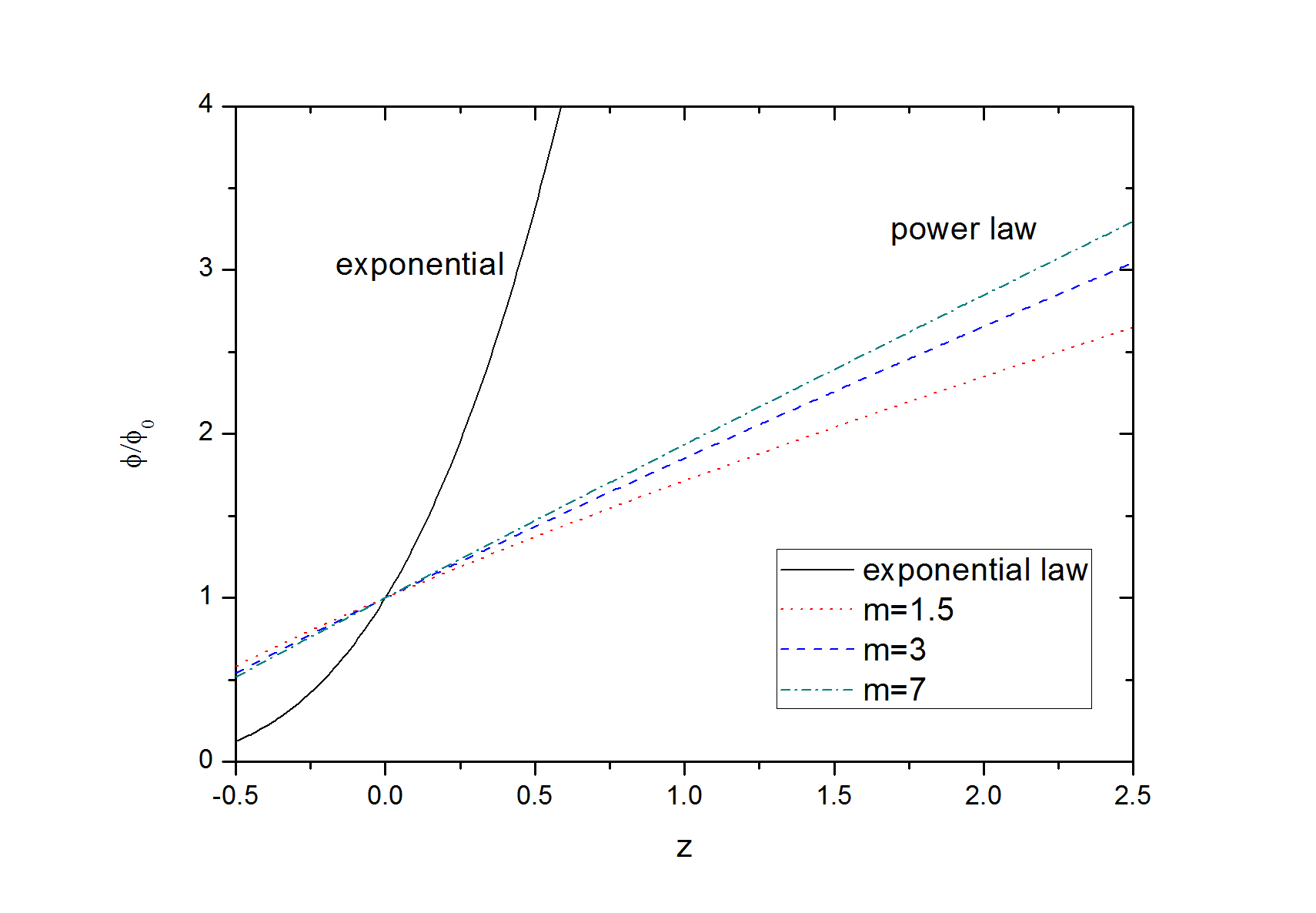}
\caption{Evolution of BD scalar field. BD field for both the exponential and power law models are shown .  For power law model, three representative values of the exponent $m$ are considered.}
\end{center}
\end{figure}

The rest energy density and pressure for the present model are,
\begin{equation}
\rho=\rho_{\Lambda}+ \rho_{\gamma} \left(\frac{\phi}{\phi_0}\right)^{1+\gamma},
\end{equation}

\begin{equation}
p=-\rho_{\Lambda}+ \gamma \rho_{\gamma} \left(\frac{\phi}{\phi_0}\right)^{1+\gamma}.
\end{equation}

The rest energy density and pressure in the model evolve with the scalar field. They decrease from higher values in the past to low values in a later period. At late times, $\rho$ dynamically evolves to become $\rho_{\Lambda}$ and the pressure $p$ reduces to $-\rho_{\Lambda}$. At late times, a negative pressure dominates the scenario and helps in the acceleration of cosmic expansion. 

Using the fact that  $\frac{\dot{\phi}}{\phi}=-3H_0$ and $\frac{\ddot{\phi}}{\phi}=9H_0^2$ we get the BD parameter as

\begin{equation}
\omega(\phi)=\omega_0+\omega_1\phi^{\gamma},
\end{equation}

where, $\omega_0=-2\left[\frac{(k^2+2k+3)}{(k+2)^2}\right]$ and $\omega_1=-\left[\frac{(\gamma+1)\rho_{\gamma}}{9H_0^2}\right] \phi_0^{-(1+\gamma)}$. It is interesting to note here that, the BD parameter has two parts: a constant $\omega_0$ and a dynamically evolving part. The constant part is decided from the anisotropic nature of the model. For an isotropic model with $k=1$, it becomes $\omega_0=-3$. The anisotropic nature of the model does not affect the evolving part of the BD parameter. The evolving part is mostly governed by the value of $\gamma$. The variable BD parameter becomes a constant for the lower limit of $\gamma$, whereas it varies linearly with the scalar field  for its upper limit. The allowed range of the BD parameter is $\omega_0+\omega_1\leq \omega(\phi)\leq\omega_0+\omega_1\phi$. The role played by the parameter $\rho^*$ is quite interesting. In the absence of this parameter, the cosmic fluid behaves as a barotropic fluid with the usual relation $p=\gamma\rho$ and $\omega_1$ turns out to be negative. Consequently, the BD parameter assumes a much higher negative value in the early phase of cosmic evolution. However, in presence of this parameter, the value of $\omega(\phi)$ is bit lifted up   because of the positive contribution from $\rho^*$. For the particular choice of $\rho^*=\left(1+\frac{1}{\gamma}\right)\rho_0$ , $\omega  _1$ vanishes and  $\omega(\phi)$ behaves as a constant $\omega_0$. In Figure -5, the functional $\omega_{BD}=\frac{\omega(\phi)-\omega_0}{\omega_1}$ is shown as a function of the scalar field for the exponential scale factor leading to a de Sitter kind of universe. The shaded area in the plot shows the allowed range of the functional $\omega_{BD}$  corresponding to the upper and lower bounds on $\gamma$. The blue curve running through the shaded area is for the representative value $\gamma=0.316$. It is obvious from the figure that, for this representative value of $\gamma$, the functional $\omega_{BD}$ increases with increase in the scalar field. At an early phase of cosmic evolution, the functional is almost constant or has a little variation with the scalar field, whereas, with the  growth of time, the rate of change in the functional becomes more rapid at late times. It can be concluded that, with the cosmic expansion, the functional $\omega_{BD}$ decreases for $\gamma >0$. The rate of decrement slows down as the value of $\gamma$ decreases from its upper bound to the lower one. For $\gamma=0$, the functional becomes a constant with a value equal to 1. However, for  $\gamma=1$, the value of $\omega$ is decided by the parameters $\rho^*, \rho_0, \phi_0$ and $H_0$. The scalar field decreases with time and therefore, for any value of $\gamma$ else than zero, the BD parameter evolves to a constant $\omega_0$ at late time of evolution. From a dimensional consistency as demanded by the Klein-Gordon wave equation (10), for $\gamma\neq 0$, the value of $\omega_0$ should be $-1.5$ which favours the anisotropic parameter $k$ to be 4. On the other hand, the average anisotropic parameter is constrained from WMAP data \cite{Campa11} to be $|\sqrt{\mathcal{A}}|=10^{-5}$ which corresponds to $m=1.0000212$ in our present model. In fact, the universe is observed to be mostly flat and isotropic and hence anisotropy in cosmic expansion must be considered as a little perturbation to the isotropic behaviour.

\begin{figure}[h!]
\begin{center}
\includegraphics[width=1\textwidth]{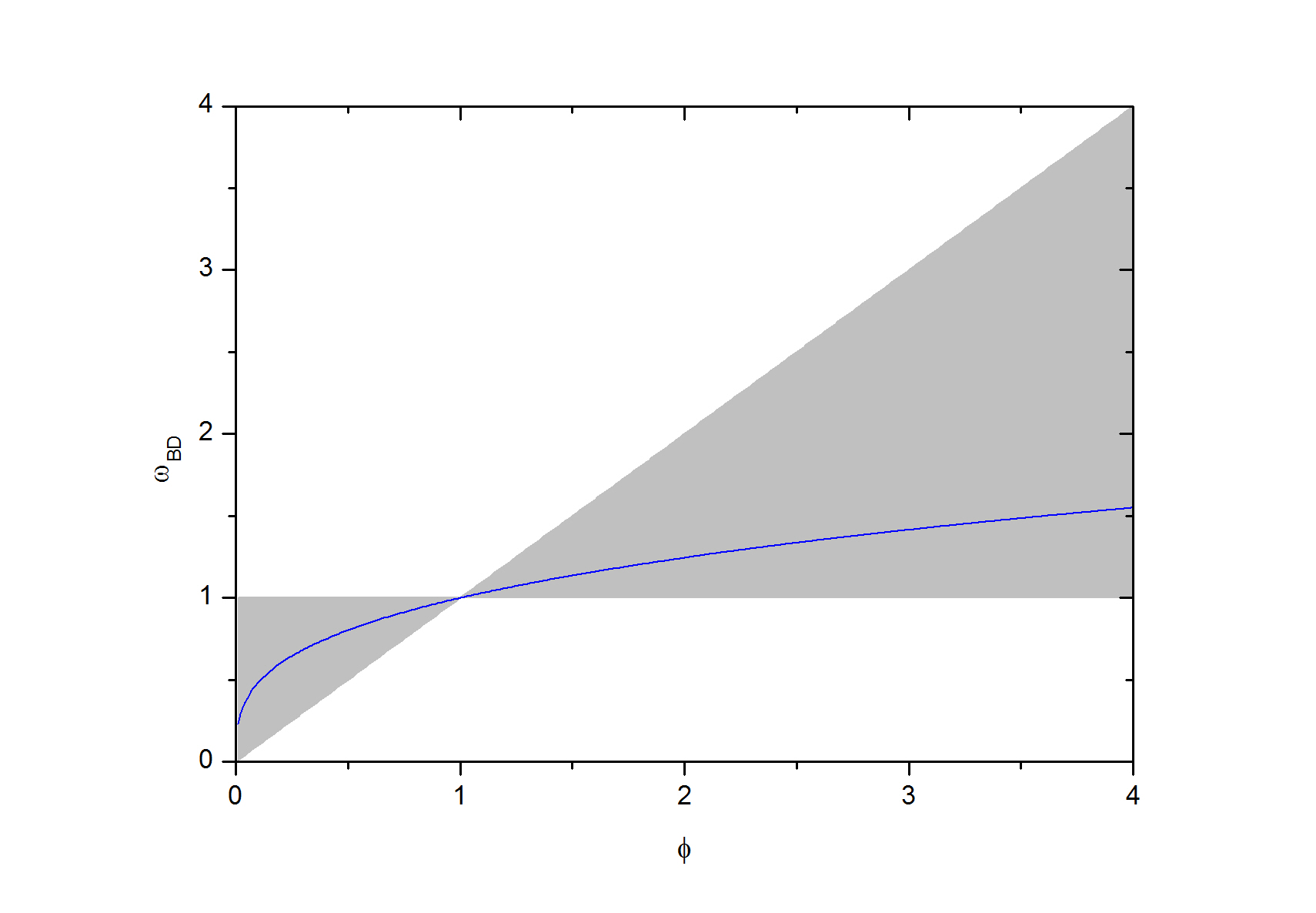}
\caption{	The functional $\omega_{BD}$, for the exponential model, as a function of scalar field. The shaded area shows the allowed range for the functional. The curve running through the shaded area is for $\gamma=0.316$.}
\end{center}
\end{figure}

\begin{figure}[h!]
\begin{center}
\includegraphics[width=1\textwidth]{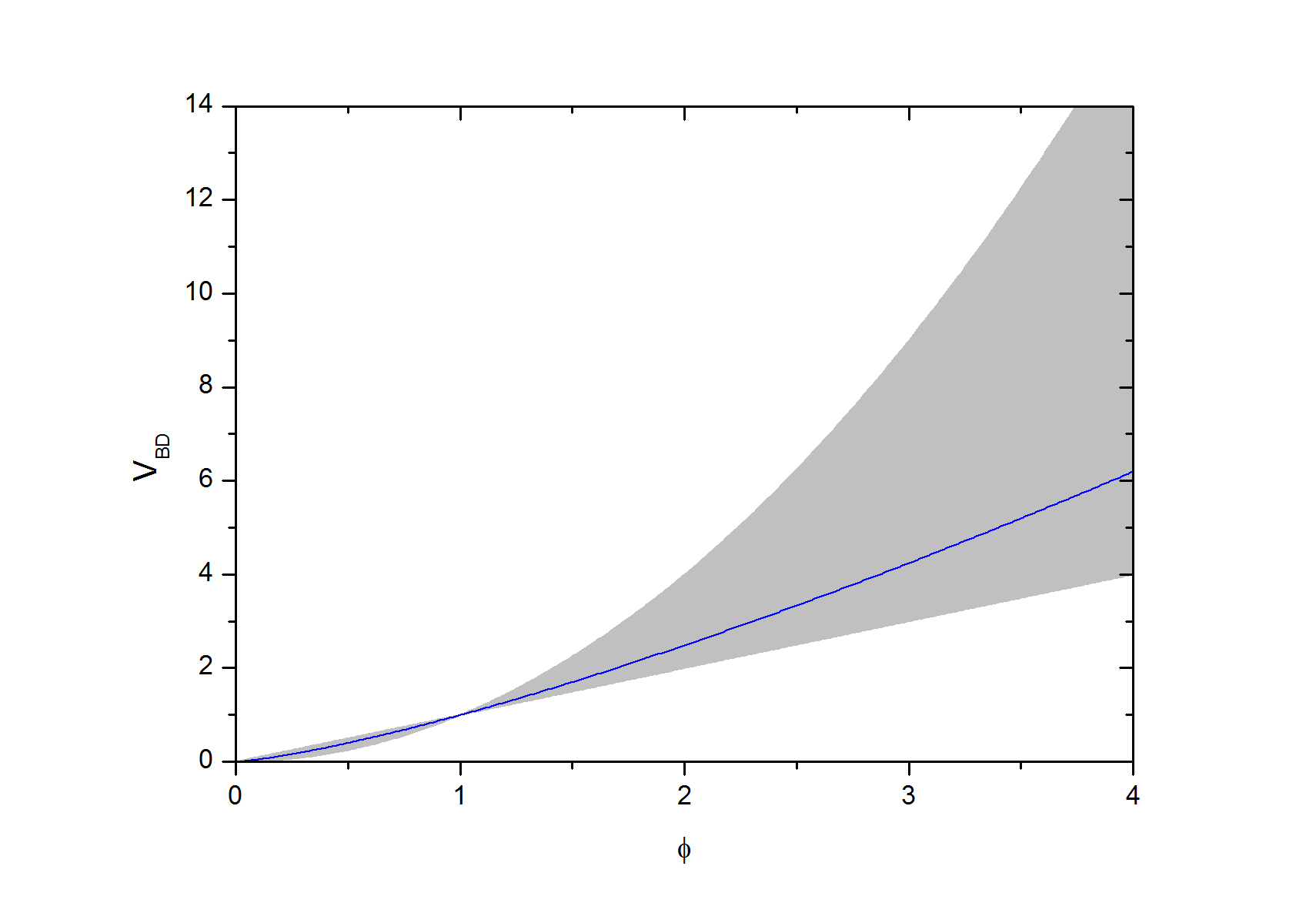}
\caption{	The functional  $V_{BD}$, for the exponential model , as a function of scalar field. The shaded area shows the allowed range for the functional. The curve running through the shaded area is for $\gamma=0.316$.}
\end{center}
\end{figure}

The self interacting potential can be expressed as
\begin{equation}
V(\phi)=V_0+V_1\phi^{1+\gamma},
\end{equation}
where,  $V_0=-2\rho_{\Lambda}$ and $V_1=-2 \rho_{\gamma}\phi_0^{-(1+\gamma)} $. The self interacting potential does not depend upon the anisotropic parameter $k$, rather it depends upon the parameters of the unified dark fluid. For a lower limit of $\gamma$, the self interacting potential varies linearly with the scalar field and for the upper limit it varies in a quadratic manner. For a particular choice of the parameter $\rho^*=\left(1+\frac{1}{\gamma}\right)\rho_0$, the BD parameter behaves like a constant with values $\omega_0=-1.5$ and the self interacting potential behaves as a constant with the value of $V(\phi)=V_0=-2\rho_0$. With the evolution of the scalar field, the self interacting potential evolves to a constant value of $-2\rho_{\Lambda}$ at late times. However, in the absence of the parameter $\rho^*$ in the dark energy eos, the potential vanishes. In other words, the presence of the parameter $\rho^*$ induces a self interacting potential even in the absence of a scalar field. The behaviour of the functional $V_{BD}=\frac{V-V_0}{V_1}$ is shown in Figure-6. The shaded area in the graph shows the allowed range of the functional $V_{BD}$. The curve running through the shaded area is for $\gamma=0.316$, where the functional $V_{BD}$ increases with the increase in the scalar field. The slope of the curve increases with the increase in $\gamma$.

\paragraph{}  The dynamics of cosmic evolution through its expansion history can be understood from the DE equation of state parameter, $\omega_{D}$. From (22) and (23), we get,

\begin{equation}
\omega_{D}=-1+\frac{1+\gamma}{1+\left(\frac{\rho_{\Lambda}}{\rho_{\gamma}}\right) \left(\frac{\phi}{\phi_0}\right)^{-(1+\gamma)}}.
\end{equation}

 The DE eos does not depend on the anisotropic nature of the model and depends on the parameters of the UDF like the self interacting potential. The DE eos, for $\gamma>0$, decreases from $\gamma$  in the quintessence region at the initial epoch to behave as a cosmological constant  with $\omega_{D}=-1$, at a later epoch when the scalar field vanishes. At a given cosmic time, the DE eos is decided by the parameters $\gamma$ and  $\rho^*$. One should note the role played by the parameter $\rho^*$. In the absence of this parameter, i.e for $\rho_{\Lambda}=0$, the DE eos is simply given by $\omega_{D}=\gamma$, which can take only positive values as decided from the constraints on the adiabatic speed of sound. But the inclusion of $\rho^*$ into the eos modifies the relation and make the DE eos a dynamic one. In other words, $\rho^*$ incorporates some negative pressure simulating the dark energy necessary for the accelerated expansion.

The time variation of Newtonian Gravitational constant is given by

\begin{equation}
\frac{\dot{G}}{G}=\frac{\dot{\phi}}{\phi}=-3H_0.
\end{equation}

Since, in the present model, the Hubble parameter is assumed to be a constant quantity through out the cosmic evolution,obviously,  $\frac{\dot{G}}{G}$  comes out to be a constant and its value can be calculated in a straightforward manner. The observational data from $H(z)$ and Supernovae Ia  constrained the Hubble parameter as $H_0=68.93^{0.53}_{-0.52} km s^{-1} Mpc^{-1}$ \cite{Kumar12} and accordingly the time variation of $G$ can be calculated from the present model.

\section{Power law Model}
In case of power law expansion with the Hubble parameter behaving as $H=\frac{m}{t}$, $m$ being a positive constant, the average scale factor behaves as $a=\left(\frac{t}{t_0}\right)^m $. The scale factors along the longitudinal and transverse directions read as $A=\left(\frac{t}{t_0}\right)^{\left(\frac{3mk}{k+2}\right)}$ and $B=\left(\frac{t}{t_0}\right)^{\left(\frac{3m}{k+2}\right)}$. Cosmologies with power law scale factor are widely discussed in literature \cite{Kumar12,Gehlaut03, Dev02, Sethi05, Batra99, Batra00, Kaplinghat99, Singh14}. The success of the power law model lies with the fact that models with $m\ge 1$ do not encounter the horizon problem and do not witness flatness problem. In Ref. \cite{Kumar12}, from the analysis of observational constraints from $H(z)$ and SNIa data, Kumar has shown that, power law cosmology is viable in the description of the acceleration of the present day universe even though it fails to produce primordial nucleosynthesis.

The deceleration parameter for this model is  $q=\frac{1}{m}-1$.  In order to be in the safe zone for accelerated expansion, the predicted deceleration parameter should be negative and that can be achieved only if $ m>1$. In terms of the deceleration parameter, the parameter $m$ can be expressed  as $m=\frac{1}{1+q}$. Considering the observational constraints from Ref. \cite{Rapetti07}, we put the constraints on $m$ to be $3.03 \le m \le 20$. Corresponding to constraints from Ref. \cite{Kumar12}, $m$ can be constrained in the range $1.4085 \le m \le 1.6393$. The jerk parameter is calculated to be $j=\frac{(m-1)(m-2)}{m}$  and  can be constrained in the range  $ 0.69 \le j \le 17.1$ \cite{Rapetti07} and $-0.1716 \le j \le -0.1407$ \cite{Kumar12}. It is worth to mention here that, the exact determination of the jerk parameter involves the observation of high-z supernovae which is a tough task. Therefore, current observational data have not yet been able to pin down the range or sign of the jerk parameter. The directional Hubble rates for this model are $H_1=\left(\frac{3mk}{k+2}\right)\frac{1}{t}$ and $H_2=\left(\frac{3m}{k+2}\right)\frac{1}{t}$. Consequently the directional deceleration parameters along different spatial directions are obtained using the relation $q_i=-1+\frac{d}{dt}\left(\frac{1}{H_i}\right)$ as $q_x=\frac{k+2}{3mk}-1$ and $q_y=q_z=\frac{k+2}{3m}-1$. The mean deceleration parameter $q$ is obtained from the directional deceleration parameters as $q=\frac{1}{3} (q_x+ q_y+ q_z)$. The directional deceleration parameters are also independent of time. For isotropic model, $k=1$ and the directional deceleration parameters all reduce to $q_x= q_y= q_z= \frac{1}{m}-1$  and become equal to the mean $q$. 
 
\paragraph{}The scalar field for this model becomes
\begin{equation}
\phi=\phi_0\left(\frac{t}{t_0}\right)^{1-3m}.
\end{equation}
In terms of the scale factor $\phi=\phi_0 \left(a\right)^{\frac{1-3m}{m}}$ and in terms of redshift $\phi=\phi_0 (1+z)^{\frac{3m-1}{3m}}$. It is obvious from (28) that, the scalar field decreases with expansion of the universe and vanishes at large cosmic time. The behaviour of the scalar field is only decided by the single parameter $m$ or more specifically the constant negative deceleration parameter. The scalar field is independent of the anisotropic parameter $k$. In Fig.-4, the scalar field for the model is shown as a function of redshift. In the figure we have considered three representative value of the exponent $m$ namely $1.5, 3$ and $7$ which are within the allowed range as calculated from the observational data for deceleration parameter. It is amply clear from the figure that, a model with higher value of $m$ has a higher scalar field in the past whereas it has a low value of scalar field in future. Also, the variation of scalar field with $m$ at early time is much exemplified than that at late times of evolution.

\paragraph{}The energy density and pressure for this model with power law expansion read as
\begin{equation}
\rho=\rho_{\Lambda}+ \rho_{\gamma}\left(\frac{\phi}{\phi_0}\right)^{\frac{3m(1+\gamma)}{3m-1}},
\end{equation}
and
\begin{equation}
p=-\rho_{\Lambda}+ \gamma \rho_{\gamma} \left(\frac{\phi}{\phi_0}\right)^{\frac{3m(1+\gamma)}{3m-1}}.
\end{equation}

Just like the previous model, the energy density and pressure evolve with the scalar field from large values at the initial epoch to respectively become $\rho_{\Lambda}$ and $-\rho_{\Lambda}$ at large cosmic time. 
\paragraph{}
The variable BD parameter can be expressed as 

\begin{equation}
\omega(\phi)=\omega_{0p}+\omega_{1p}\phi ^{\left(\frac {3\gamma m-1}{3m-1}\right)},
\end{equation}

where, $\omega_{0p}=\frac {3m[ (k+2)(k-3mk+2)-6m(1-k)]}{(1-3m)^2 (k+2)^2 }$ and $\omega_{1p}=-\frac{(\gamma+1)\rho_{\gamma}}{(1-3m)^2}t_0^2\phi_0^{-\frac{3\gamma m-2}{3m-1}}$. We have used the fact $\frac{\dot{\phi}}{\phi}=\frac{1-3m}{t}$  and $\frac{\ddot{\phi}}{\phi}=\frac{3m(3m-1)}{t^2}$ to get above relation (31) from (14). It is interesting to note that, the BD parameter is a function of the scalar field even in the lower limit of $\gamma$, in which it decreases with the scalar field. In other words, the BD parameter assumes lower values in the past and larger values in the late time of cosmic evolution. If we consider the upper bound of $\gamma$, the BD parameter evolves linearly with the scalar field. The anisotropic nature of the model affects only the constant part of BD parameter. The behaviour of the evolving part  is governed by the parameters of the UDF and the exponent $m$.  In Figure-7, the functional $\omega_{BD}=\frac{\omega-\omega_{0p}}{\omega_{1p}}$ is plotted as a function of the BD field. The shaded area shows the allowed range. In order to get a general behaviour, we have shown the functional for a representative value $\gamma=0.316$. For the upper bound of $\gamma$, the functional linearly behaves with the BD field. In order to calculate the lower bound for the functional we have used a reasonable value of the exponent $m=1.5$ which lies within the observational limits corresponding to a more recent data. Just like the previous model, the functional $\omega_{BD}$ for the representative value of $\gamma$,varies slowly with BD field at an early epoch and varies rapidly at late time of evolution. $\rho^*$ has a significant role upon the behaviour of the BD parameter. For the particular choice  $\rho^*=\left(1+\frac{1}{\gamma}\right)\rho_0$, it behaves as a pure constant which can be equated to $-1.5$, from dimensional consistency of the Klein-Gordon wave equation.

\begin{figure}[h!]
\begin{center}
\includegraphics[width=1\textwidth]{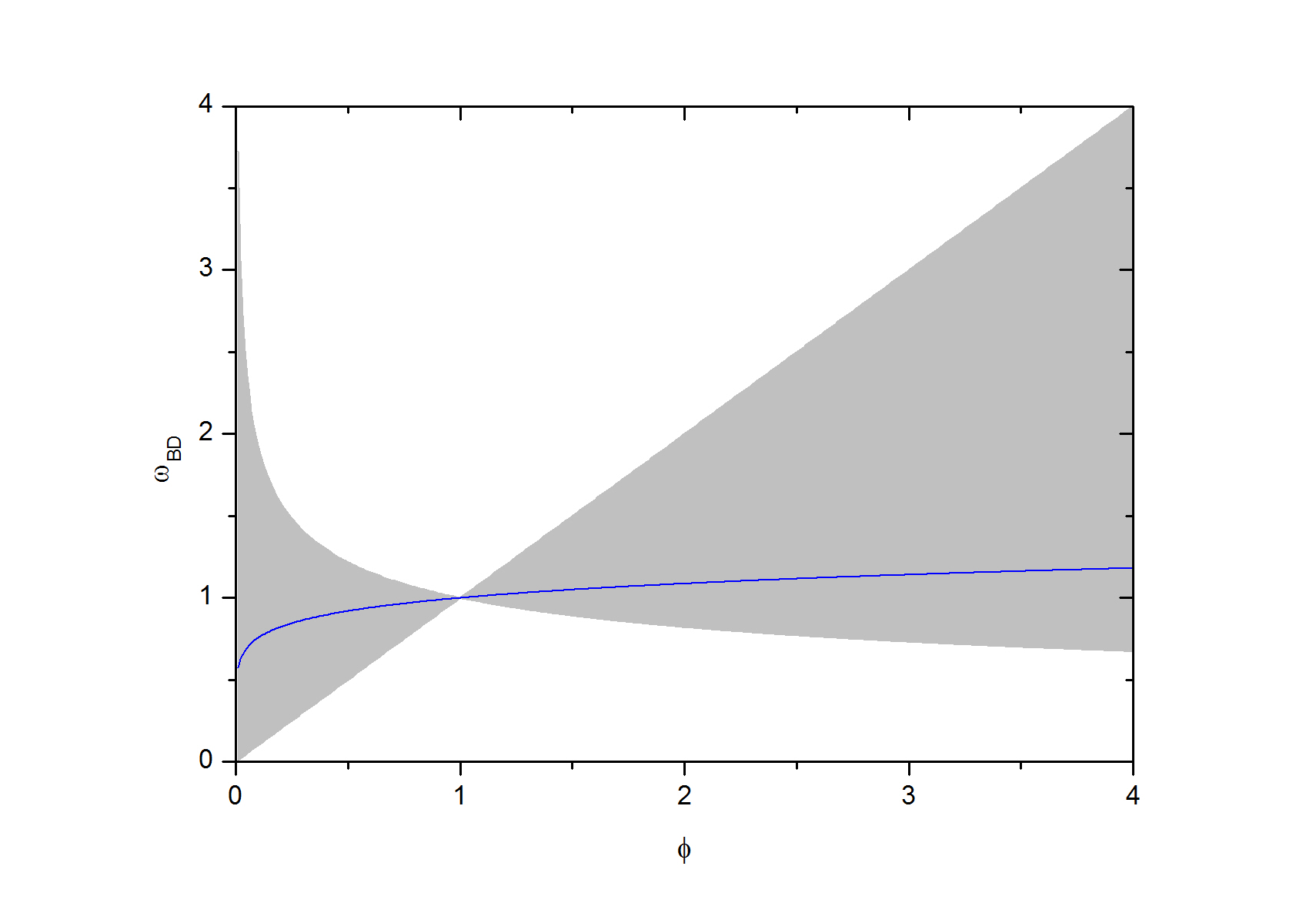}
\caption{The functional $\omega_{BD}$, for the power law model, as a function of scalar field. The shaded area shows the allowed range for the functional. The curve running through the shaded area is for $\gamma=0.316$.}
\end{center}
\end{figure}

\begin{figure}[h!]
\begin{center}
\includegraphics[width=1\textwidth]{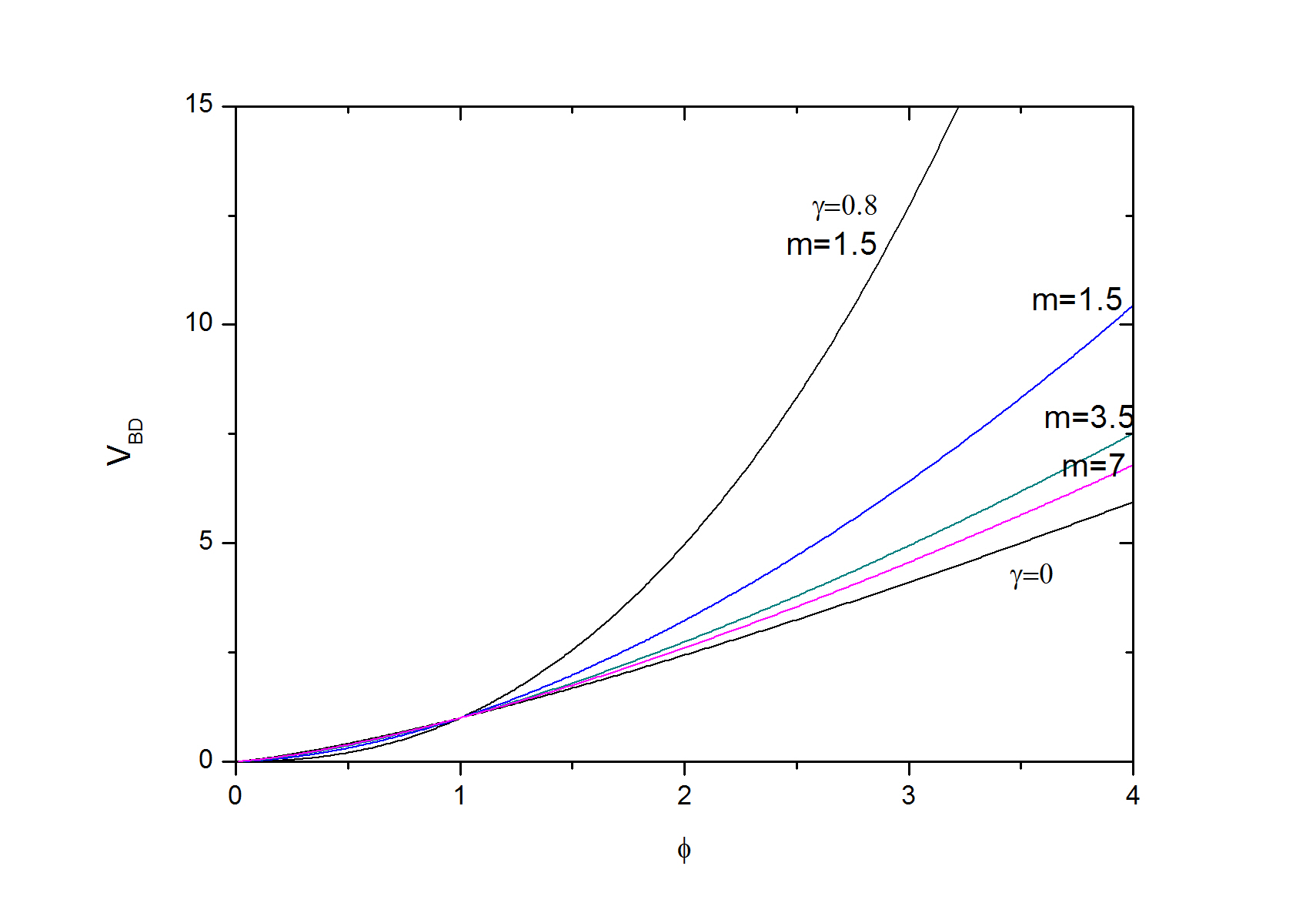}
\caption{The functional $V_{BD}$ , for the power law model , as a function of scalar field. The upper curve is for $\gamma=0.8$ and $m=1.5$. The lower curve shows the lower bound with $\gamma=0$.  The three curves  in the middle are for three different values of the exponent $m$ with  $\gamma=0.316$.}
\end{center}
\end{figure}

The self interacting potential for this model is given by
\begin{equation}
V(\phi)=V_0+V_{1p}\phi^{\frac{3m(1+\gamma)}{3m-1}},
\end{equation}
where, 
\begin{equation}
V_{1p}=(\gamma-1) \rho_{\gamma}\phi_0^{\left(\frac{3m(1+\gamma)}{1-3m} \right)}.
\end{equation}
Since, $m>1$, the self interacting potential increases with the increase in the scalar field. Like the previous model, the scalar field does not depend on the anisotropic exponent $k$ and it depends on the parameters of the unified dark fluid. For a choice of  $\rho^*=\left(1+\frac{1}{\gamma}\right)\rho_0$ or  $\gamma=1$, the self interacting potential becomes independent of the scalar field and equals to $-2\rho_{\Lambda}$. This is the same value the potential assumes at a later epoch. In other words, there is an induced self interacting potential in the absence of the scalar field, because of the parameter $\rho^*$. In Figure-8, we have shown the functional $V_{BD}=\frac{V-V_{0}}{V_{1p}}$ as a function of BD field. In this figure we can not set up the upper bound since $V_{1p}$ vanishes for $\gamma=1$. However, a curve for $\gamma=0.8$  with $m=1.5$ is shown in the figure to get an idea. The curves for $\gamma=0.316$ are shown for three different values of $m$ e.g. $m=1.5,3.5$ and $7$. The functional $V_{BD}$ decreases with the decrease in the field and at late times of evolution, it vanishes. For a given value of $\gamma$, the functional decreases with the increase in $m$ at early epochs whereas it increases at late times. However, the rate of increment at late times is less as compared to the rate of decrement at early phase.

\paragraph{}The DE equation of state $\omega_{D}$ can be calculated from (29) and (30) as
\begin{equation}
\omega_{D}=-1+\frac{1+\gamma}{1+\left(\frac{\rho_{\Lambda}}{\rho_{\gamma}}\right) \left(\frac{\phi}{\phi_0}\right)^{\frac{3m(1+\gamma)}{1-3m}}}.
\end{equation}
The DE eos decreases from $\gamma$ in the beginning to behave like a cosmological constant  with $\omega_{D}=-1$ at a late epoch of cosmic evolution. In the absence of the parameter $\rho^*$, the DE eos is a constant quantity i.e. $\gamma$. The presence of this parameter makes the DE eos an evolving one. The anisotropic nature of the model does not affect $\omega_{D}$. However, the DE eos is controlled by the choice of the exponent $m$ which is decided by the observational constraints on the deceleration parameter and the jerk parameter.

The time variation of Newtonian Gravitational constant for this power law model is

\begin{equation}
\frac{\dot{G}}{G}=\frac{1-3m}{t}.
\end{equation}

Here, $\frac{\dot{\phi}}{\phi}=\frac{1-3m}{t}$ inversely varies with time. The value of $m$ for the  present is constrained from the observational data \cite{Kumar12} and consequently the time variation of $G$ can be predicted to be in  the range $-3.918 < \frac{\dot{G}}{G} t < -3.226$.

\section{Conclusion}
In the present work, we have constructed some cosmological models mimicking the late time cosmic acceleration in the frame work of generalized Brans-Dicke scalar tensor theory of gravitation for a plane symmetric universe. The cosmic fluid is considered to be a dark fluid described by two parameter affine equation of state.  The shear scalar is considered to be proportional to scalar expansion which simulates a linear relationship among the directional Hubble rates incorporating anisotropy in expansion rates along different spatial directions. In general relativity, such an assumption does not provide accelerating model. However, in the frame work of generalised BD theory with evolving scalar field, it is possible to get accelerated phase of expansion with such assumption. Considering a constant deceleration parameter at a late time of evolution of the universe, we have considered two kinds of volume expansion namely, the power law expansion and the exponential law of expansion. Moreover, we have shown that, a constant deceleration parameter leads to a power law in the BD scalar field. The presence of the extra term in the barotropic fluid eos, makes the dark energy eos an evolving one. The DE eos evolves from a positive constant quantity equal to the adiabatic speed of sound in the beginning to  behave like a cosmological constant at a later epoch of cosmic evolution. The scalar field is found to decrease with the cosmic expansion. The self interacting potential increases with the increase in scalar field. In an initial epoch, the self interacting potential is having a large value and decreases with time to have a constant value decided by the equation of state parameter at a later epoch. The anisotropic nature of the model does not affect the behaviour of the scalar field and the self interacting potential. However, the non-evolving part of the dynamic BD parameter is affected by the introduction of an anisotropy in the expansion rates.

\section{Acknowledgement}
BM acknowledges University Grants Commission, New Delhi, India for financial support to carry out the Minor Research Project [F.No-42-1001/2013(SR)]. SKT likes to thank Saha Institute of Nuclear Physics, Kolkata, India for providing necessary facilities where a part of this work is done.

\end{document}